\documentclass[12pt]{article}
\usepackage{cite}
\usepackage{wrapfig}
\usepackage{graphicx}
\usepackage{amssymb}
\usepackage{amsfonts}
\usepackage{amsmath}
\usepackage{longtable}
\usepackage{rotating}
\usepackage{lscape}
\usepackage{epsfig}
\usepackage{multirow}
 \usepackage[utf8]{inputenc}

\usepackage[dvipsnames]{xcolor}
 \usepackage{hyperref}

\begin{document}

\date{}

\title{\bf\ Neutrino Mass in Effective Field Theory}
 \maketitle

\setcounter{footnote}{0}

  \begin{center}

 {\it \large \bf A.V.\,Borisov$^{a,}$\footnote{E-mail: borisov@phys.msu.ru} and
\bf A.P.\,Isaev$^{a,b,}$\footnote{E-mail: isaevap@theor.jinr.ru}}

 {$^{a}$\,M. V. Lomonosov Moscow State University,
 Faculty of Physics, Moscow, Russia}

 {$^{b}$\,Joint Institute for Nuclear Research, \\ N.N.~Bogolubov Laboratory of Theoretical Physics, Dubna}

 \end{center}
 \vspace{1cm}

\begin{center}
{\bf Abstract.}
\end{center}
In this review,
the seesaw mechanism for generating the mass of active light neutrinos (both Majorana and Dirac) is considered on the basis of effective field theory. In particular, we review certain models that extend the
Standard Model by introducing heavy sterile neutrinos and
discuss the corresponding mechanisms for generating
small masses of active neutrinos.
Two Appendices briefly describe the properties of Weyl, Dirac,
 and Majorana spinors in four dimensions and the interrelations
 between such spinors. The third Appendix provides
a simple proof of the theorem on Takagi diagonaliza\-tion
 of a mass matrix for Majorana fermions.

\vspace*{6pt}

\noindent
PACS: 14.60.Pq; 14.60.St

\label{sec:intro}
\section*{Introduction}

In the minimal Standard Model (SM), neutrinos are massless left-chiral (left-handed) fermions \cite{wei} forming, together with the left-handed components of charged leptons, three doublets that are transformed according to the fundamental representation of the gauge group ${\rm SU(2)}_ {L}$.
However, the discovery of neutrino oscillations showed that neutrino masses are non-zero, but for the three active neutrinos they are very small (see review \cite{pdg}). From the data of oscillations, only two differences of squared neutrino masses (in the 3-neutrino mixing scheme) \cite{pdg} are determined, but not their absolute values, which allows us to obtain a lower limit on the largest of the three masses:
\begin{equation}
\label{m3}
m_3 > \sqrt{\Delta m_{31}^{2}} \simeq 0.05~\text{eV}.
\end{equation}
The most stringent upper limit on the sum of light neutrino masses from modern cosmological data is \cite{dival}:
\begin{equation}
\label{cosm}
\sum{m_{\nu } } \equiv \sum\limits_{i = 1}^{3} m_{i} < 0.09~\text{eV}.
\end{equation}

In the minimal SM, the masses of fermions (charged leptons and quarks) are generated due to the Yukawa interaction of the Higgs scalar doublet $\varphi$ with doublets of left-handed fermion components and right-handed fermion singlets, but neutrinos, being only left-handed, remain massless.
For the generation of Dirac neutrino masses, which already makes it possible to describe neutrino oscillations, it is sufficient to introduce right-handed compo\-nents of neutrino fields $\nu_{\alpha R}$ ($\alpha = e,\mu,\tau$) and use the same
Brout--Engler--Higgs mechanism as for charged fermions. However,
it should be emphasized that right-handed neutrinos are fundamentally different from left-handed neutrinos and right-handed charged fermions. Namely, $\nu_{\alpha R}$ are sterile (unlike active $\nu_{\alpha L}$), i.e. they do not participate in electroweak (and, of course, strong) interactions, since their weak isospin and hypercharge are zero (fields $\nu_{\alpha R}$
 are singlets of the gauge group $\text{SU(3)}_{c}\times \text {SU(2)}_{L}\times\text{U(1)}_{Y}$).

The Lagrangian of the interaction generating Dirac neutrino masses has the form
(we follow \cite{giu})
\begin{equation}
\label{Y}
{\cal L}_{\nu H} = - y_{\alpha\beta}^{\nu}{\bar L}_{\alpha L}^{\prime}{\tilde\varphi} \nu_{\beta R}^{\, \prime} + \text{H.c.}
\end{equation}
Here $y_{\alpha\beta}^{\nu}$ are complex Yukawa coupling constants,
indices $\alpha,\beta = e,\mu,\tau$ enumerate lepton generations, and
${\tilde\varphi} = i\sigma_{2}\varphi^{+T}$
(where $\sigma_2$ is the Pauli matrix, see (\ref{sigma1}))
 is the doublet charge conjugated
to the Higgs doublet
\begin{equation}
\label{f}
\varphi = \begin{pmatrix} \phi^{+}\\ \phi^{0} \end{pmatrix} \, ,
\;\;\;\; L_{\alpha L}^{\prime} =
\begin{pmatrix} \nu_{\alpha L}^{\, \prime}\\
\ell_{\alpha L}^{\, \prime}\end{pmatrix},
\end{equation}
with primes indicating fields with indefinite masses.

After spontaneous breaking of gauge symmetry,
 a nonzero vacuum expectation
 value of the Higgs field arises, so that in the unitary gauge
 we have
 \begin{equation}
\label{vac}
\langle 0|\tilde\varphi|0\rangle = (v/\sqrt{2},0)^{T}\, , \;\;\;\;
 v =(\sqrt{2}G_{\text F})^{- 1/2} \simeq 246~\text{GeV},
\end{equation}
and Lagrangian (\ref{Y}) takes the form (in matrix notation):
\begin{equation}
\label{LH}
{\cal L}_{\ell H} = - \frac{1}{\sqrt{2}}(v + H){\bar \nu}_{L}^{\prime}Y^{\nu}\nu_{R}^{\prime} + \text{H.c.},
\end{equation}
where $H$ is the scalar Higgs field, $Y^{\nu} = (y_{\alpha\beta}^{\nu})$ denotes the matrix of the Yukawa couplings,
$\nu_{P}^{\prime} = (\nu_{e P}^{\prime}, \nu_{\mu P}^{\prime},\nu_{\tau P}^{\prime})^{T}$, $P = L, R$ (see eq. (\ref{LR}) in Appendix A).

After the bi-unitary diagonalization of the matrix $Y^{\nu}$:
\begin{equation}
\label{bud}
\frac{v}{\sqrt{2}}(V_{L}^{\nu})^{+}Y^{\ell}V_{R}^{\nu} =
{\frac{v}{\sqrt{2}} \text{diag}(y_1,y_2,y_3)
 = } \text{diag}(m_1,m_2,m_3) \equiv M^{\nu} ,
\end{equation}
we obtain the Lagrangian
\begin{equation}
\label{LHp}
{\cal L}_{\nu H} = - \left(1 +\frac{H}{v}\right)\left({\bar n}_{L}M^{\nu}n_{R}+ \text{H.c.}\right) =  - \left(1 +\frac{H}{v}\right)
\sum\limits_{k = 1}^{3} m_{k}{\bar \nu}_{k}\nu_{k}
\end{equation}
in terms of physical fields (with definite masses)
\begin{equation}
\label{nLR}
n_L = (V_{L}^{\nu})^{+}\nu_L^{\, \prime} = (\nu_{1L},\nu_{2L},\nu_{3L})^{T}
\, , \;\;\;\;
n_R = (V_{R}^{\nu})^{+}\nu_R^{\, \prime}  = (\nu_{1R},\nu_{2R},\nu_{3R})^{T}.
\end{equation}
The Lagrangian (\ref{LHp}) includes the mass terms (here
$\nu_{k} = \nu_{k L} + \nu_{k R}$ are
four-component Dirac fields) and the term of interaction of massive neutrinos with the Higgs boson, and what is more, the neutrino masses expressed through the Yukawa couplings $y_{k}$ and the vacuum condensate $v$ (see (\ref{bud})):
\begin{equation}
\label{mk}
m_{k} = y_{k}\frac{v}{\sqrt{2}}.
\end{equation}

The lepton weak charged current, which describes the interaction with $W$ bosons, includes fields of the left-handed neutrinos with definite flavors
$\nu_{L} = (\nu_{eL}, \nu_{\mu L}, \nu_{\tau L})^{T}$ which are
superpositions of the left-handed neutrino fields with definite masses
(see (\ref{nLR})):
\begin{equation}
\label{cc}
j_{\lambda}^{(-)} = {\bar \ell}_{L}^{\, \prime}
\gamma_{\lambda}\nu_{L}^{\, \prime} = {\bar \ell}_{L}(V_{L}^{\ell})^{+}\gamma_{\lambda}V_{L}^{\nu}n_{L}
={\bar \ell}_{L}\gamma_{\lambda}\nu_{L} =  \sum\limits_{\alpha = e,\mu,\tau}^{}{\bar \ell}_{\alpha L}\gamma_{\lambda}\nu_{\alpha L}.
\end{equation}
Here
\begin{equation}
\label{fl}
\nu_{L} = U\, n_{L} \; , \;\;\; \nu_{\alpha L} =
\sum\limits_{k = 1}^{3}\, U_{\alpha k}\, \nu_{kL} \; ,
\end{equation}
and $U = {||U_{\alpha k}||} =
(V_{L}^{\ell})^{+}V_{L}^{\nu}$ is the unitary
Pontecorvo--Maki--Nakagawa--Sakata (PMNS) matrix of lepton mixing,
for which, in the case of Dirac neutrinos, the standard
parametrization\footnote{In the case of the Majorana neutrinos, the
 PMNS-matrix (\ref{Mckm2}) is modified \cite{pdg}:
 $U \to U \cdot {\rm diag}(e^{i\eta_{1}}, e^{i\eta_{2}},1)$
 and depends on three $CP$-violating  phases
 $\delta,\eta_1,\eta_2$.} is
  \begin{equation}
  \label{Mckm2}
 U = \left(\!\! \begin{array}{ccc}
 1 & 0 & 0 \\
 0 & c_{23} & s_{23} \\
  0 & - s_{23} & c_{23}
 \end{array} \!\!\right)
 \left(\!\! \begin{array}{ccc}
 c_{13} & 0 & s_{13} \, e^{-i\delta} \\
 0 & 1 &  0 \\
 - s_{13} \, e^{i\delta} & 0 & c_{13}
 \end{array} \!\!\right)
 \left(\!\! \begin{array}{ccc}
 c_{12} & s_{12} & 0  \\
 - s_{12} & c_{12} & 0  \\
 0  &  0 & 1
 \end{array} \!\!\right) \; ,
  \end{equation}
  where we use the concise notation:
$s_{ij} := \sin \theta_{ij}$, $\;c_{ij} := \cos \theta_{ij}$.

We emphasize that the Standard Model determines only the general structure of the matrix (\ref{Mckm2}) but does not predict the numerical values of its parameters, which are determined from experimental data on neutrino oscillations
(see \cite{pdg,nufit,olshev,sal}). For definiteness, we present data from \cite{sal} correspon\-ding to the best fit and two options of the hierarchy
 of the neutrino mass spectrum (see \cite{pdg,giu}), normal and inverted  (indicated in parentheses):
 \begin{equation*}
\begin{gathered}
  s_{12}^2/10^{-1} = 3.18 \pm 0.16 \; (3.18 \pm 0.16), \\[0.1cm]
  s_{23}^2/10^{-1} = 5.74 \pm 0.14 \; (5.78^{+0.10}_{-0.17}), \\[0.1cm]
 s_{13}^2/10^{-2}  = 2.200^{+0.069}_{-0.062} \; (2.225^{+0.064}_{-0.070}),
 \\[0.1cm]
 \delta/\pi = 1.08^{+0.13}_{-0.12} \; (1.58^{+0.15}_{-0.16}).
\end{gathered}
\end{equation*}
As it can be seen, a reliable value of the $CP$-violating phase $\delta$ cannot yet be extracted from modern experimental data.

Equation (\ref{mk}) is applicable for any Dirac fermions. From this it follows that the Yukawa coupling constant $y_{f}$ of the fermion with the Higgs boson
increases with increasing mass of the fermion, and experimental data \cite{pdg} demonstrate a huge hierarchy of the mass spectrum of fundamental fermions and corresponding $y_{f}$. So taking into account (\ref{vac}), for the neutrino (we choose its mass $m_{\nu} \simeq 0.05~\text{eV}$, see (\ref{m3})), electron and $t $-quark we obtain:
\begin{equation}
\label{hy}
y_{\nu} \simeq 3\times 10^{- 13} \, , \;\;\;\; y_{e} \simeq 3\times 10^{- 6}
 \, , \;\;\;\; y_{t} \simeq 1.
\end{equation}

This hierarchy is one of the fundamental problems of
elementary particle physics, which cannot be solved within the framework of the Standard Model and requires its extension \cite{em,nag,lang,boos}.

To study the effects of new physics not described by the Standard Model, the concept of effective field theory (EFT) \cite{pet,mei,boos2,falk} developed by S. Weinberg \cite{wei2,wei3} is used.

Under the assumption that the energy scale
  $\Lambda$ of new physics is significantly larger than the characteristic SM scale $v$ (see (\ref{vac})) the EFT Lagrangian is represented as
\begin{equation}
\label{eff}
{\cal L}_{\text{eff}} = {\cal L}_{\text{SM}} + \sum\limits_{n > 4}^{}{\Lambda^{- n}}\sum\limits_{k}^{}C_{k}^{(n)}{\cal O}_{k}^{(n)}.
\end{equation}
Here ${\cal L}_{\text{SM}}$ is the SM Lagrangian, and the remaining terms describe the effects of new physics and
include composite operators ${\cal O}_{k}^{(n)}$ with mass dimension
$n = 5, 6,\dots$; $C_{k}^{(n)}$ are the numerical dimensionless
parameters.
The operators ${\cal O}_{k}^{(n)}$ are invariant under the SM gauge group $\text{SU(3)}_{c}\times \text{SU(2)}_{L}\times\text{U(1)}_{Y}$
and are composed only of the SM fields.
The coefficients $C_{k}^{(n)}$ are determined from experimental data or, if the Lagrangian of a particular theory extending the SM is known,
$C_{k}^{(n)}$
are expressed in terms of coupling constants and masses of new (heavy) particles by matching the amplitudes of physical processes obtained on the basis of the two specified Lagrangians (in the region of relatively low energies $E \ll \Lambda$, where the
effective Lagrangian (\ref{eff}) is applicable).
The Lagrangian (\ref{eff}) can also be defined as the
Lagrangian for the effective action obtained from the generating functional of the extended theory by integration over
``heavy'' fields \cite{pet,mei,boos2,falk}.

There is a unique set of operators ${\cal O}^{(5)}$ of dimension 5 composed of the SM fields and possessing gauge symmetry \cite{wei4}:
\begin{equation}
\label{O5}
{\cal O}^{(5)} = z_{\alpha\beta}\left({\bar L}_{\alpha L}^{\prime}{\tilde\varphi}\right)\left({\tilde\varphi}^{T} L_{\beta L}^{\prime c}\right) + \text{H.c.}
\end{equation}
Here $z_{\alpha\beta} = z_{\beta\alpha}$ is a set of (complex)
dimensionless constants, $L_{\beta L}^{\prime c} =
C{\bar L}_{\beta L}^{\prime T}$ is the charge conjugated doublet and $C$ is the charge conjugation operator (see Eq. (\ref{chc}) in Appendix A).
We stress that the operator ${\cal O}^{(5)}$, which is absent in the SM Lagrangian, does not preserve the total lepton number, changing it by two units. After spontaneous symmetry breaking,
this operator generates a mass term for neutrinos
\begin{equation}
\label{nuM}
{\cal L}_{\nu M} = - \frac{1}{2}M_{\alpha\beta}^{\prime}{\bar \nu}_{\alpha L}^{\prime}\nu_{\beta L}^{\prime c} + \text{H.c.} \, , \;\;\;\;
 M_{\alpha\beta}^{\prime} = z_{\alpha\beta}\frac{v^2}{\Lambda}.
\end{equation}
The symmetric (complex) mass matrix $M^{\prime}$ is transformed to the diagonal form by using the unitary matrix $V$ (see \cite{giu}, as well as the comment on
Eq. (\ref{M}) below):
\begin{equation}
\label{dM}
V^{T}M^{\prime}V = M = \text{diag}(m_1,m_2,m_3),
\end{equation}
where $m_k$ are positive numbers, and the initial left-handed flavor fields are represented in the form of left-handed field components
with definite masses:
\begin{equation}
\label{nL}
\nu_{L}^{\prime} = V \, n_{L} \, , \;\;\;\;  n_{L} =
(\nu_{1L},\, \nu_{2L},\, \nu_{3L})^{T}.
\end{equation}
Making use of Eqs. (\ref{dM}) and (\ref{nL}), we reduce
the mass term (\ref{nuM}) to the diagonal form
\begin{equation}
\label{Maj}
{\cal L}_{\nu M} = - \frac{1}{2}\left({\bar n}_{L}Mn_{L}^{c} +  {\bar n}_{L}^{c}Mn_L\right) =
- \frac{1}{2}\sum\limits_{k = 1}^{3}
m_{k} \; {\bar \nu}_{k}\nu_{k} \; ,
\end{equation}
where
\begin{equation}
\label{nuMaj}
\nu_{k} = \nu_{kL} + \nu_{kL}^{c} = \nu_{k}^{c}.
\end{equation}
Thus, massive neutrinos turn out to be Majorana particles
(see Appendix A) coinciding with their antiparticles. As it
follows from (\ref{nuM}) and (\ref{dM}), their masses
\begin{equation}
\label{mz}
m_k = z_k\frac{v^2}{\Lambda},
\end{equation}
are significantly less than the masses of charged
leptons (see (\ref{mk})) due to the presence of the suppressing
factor $v/\Lambda$ caused by the effects of new physics.

The typical neutrino mass scale can be represented as
\begin{equation}
\label{msc}
m_\nu \sim \frac{v^2}{\Lambda} \simeq 6\times 10^{- 2} \times \left(\frac{10^{15}~\text{GeV}}{\Lambda}\right)\text{eV} \; ,
\end{equation}
where $10^{15}~\text{GeV}$ is the typical energy scale for
grand unified models.

In this paper, we review a number of models that extend the SM by introducing heavy sterile neutrinos and
discuss the corresponding mechanisms for the
generation of  small masses of active neutrinos.

\section{Seesaw Mechanism for Generating Neutrino
Masses\label{sec:seesaw}}

To explain the small masses of active neutrinos, a seesaw mechanism (SSM) of their generation was proposed, which is caused by the interaction of flavor neutrinos with heavy right-handed Majorana neutrinos \cite{min,g-m,yan,gl,moh}. There are three types of SSM
 which are classified in \cite{ma} (see also \cite{xi}).

 In this paper, we will limit ourselves to considering SSM of type I, which is based on expanding the SM by adding three heavy right-handed neutrinos (singlets of the gauge group $\text{SU(2)}_{L}$) while preserving the standard Higgs doublet. For the SSM of type II, a heavy Higgs triplet is added; for type III, a triplet of heavy left-handed fermions is added (various modifications and combinations of
 all these 3 mechanisms are also possible \cite{xi}).

All these mechanisms lead to non-conservation of the lepton number. We also note that the detailed experimental studies of Higgs boson properties, carried out after its discovery in 2012, are in good agreement with the predictions of the Standard Model: so far no signals of new physics have been detected in the Higgs sector \cite{pdg}.

We first consider the SSM of type I for a simple model of one lepton doublet $L_{L} = (\nu_{L}, e_L)^{T}$ interacting with a heavy right-handed neutrino\linebreak (singlet) $N_R$. The corresponding part of the full Lagrangian has the form
\begin{equation}
\label{NR}
{\cal L}_{\nu N} = {\bar N}_Ri\gamma^{\mu} {\partial}_{\mu}N_R - \frac{1}{2}{m_R}\left( {\bar N_R^c{N_R} + {{\bar N}_R}N_R^c} \right) - {y_\nu }\left( {{{\bar L }_L}\tilde \varphi {N_R} + {{\bar N}_R}{{\tilde \varphi }^ + }{L_L}} \right),
\end{equation}
where the mass is $m_R \gg v$ (it is assumed that it is
generated by new physics not described by the SM).
After spontaneous symmetry breaking on the $v$ scale, the mass part of the Lagrangian arises, which is a superposition of the Dirac and Majorana mass terms
\begin{equation}
\label{DM}
{\cal L}_{DM} = - {m_D}\left( {{{\bar \nu }_L}N + \bar N{\nu _L}} \right) - \frac{1}{2}{m_R}{\bar N}N,
\end{equation}
where we introduced the notation for the Dirac mass
\begin{equation}
\label{mD}
 m_D = y_{\nu}\frac{v}{\sqrt{2}},
\end{equation}
 and $ N = N_R + N_{R}^{c} = N^{c}$ is a Majorana neutrino field. Next, we note that, in view of
 $\nu _L^c \equiv {({\nu _L})^c} = {({\nu ^c})_R}$, we have
\begin{equation}
\label{rel}
\begin{gathered}
{\nu _L} = \frac{{1 - {\gamma ^5}}}{2}\chi \,  ,
\quad \quad \chi  = {\nu _L} + \nu _L^c = {\chi ^c},\quad \\
{{\bar \nu }_L}N + \bar N{\nu _L} = \frac{1}{2}\left( {\bar\chi  N + \bar N\chi } \right) + \frac{1}{2}\left( {\bar \chi {\gamma ^5}N - \bar N{\gamma ^5}\chi } \right),
\end{gathered}
\end{equation}
and taking into account the relations
(\ref{chc}), (\ref{bpc}) and (\ref{g5}) (see  Appendix A)
\[
\bar N{\gamma ^5}\chi  =  - {\chi ^T}({\gamma ^{5})^{T}}{{\bar N}^T} = ( - {\chi ^T}{C^{ - 1}})(C{(\gamma ^{5})^{T}}{C^{ - 1}})(C{{\bar N}^T}) = {{\bar \chi }^c}{\gamma ^5}{N^c} = \bar \chi {\gamma ^5}N
\]
we represent (\ref{DM}) in the matrix form through the Majorana fields $N$ and $\chi$:
\begin{equation}
\label{MM}
{\cal L}_{DM} = - \frac{1}{2}(\bar\chi,\bar N)\begin{pmatrix} 0&m_D\\
m_D&m_R\end{pmatrix}\begin{pmatrix}\chi\\N\end{pmatrix}.
\end{equation}
After diagonalizing the mass matrix in (\ref{MM}),
\begin{equation}
\label{diag}
\begin{gathered}
 U^{T}\begin{pmatrix} 0&m_D\\m_D&m_R\end{pmatrix} U = \begin{pmatrix} m_{1}&0\\
0&m_{2}\end{pmatrix},\quad U = RP, \\
R = \begin{pmatrix} \cos\theta& \sin\theta\\
-\sin\theta&\cos\theta \end{pmatrix},\quad P = \begin{pmatrix} - i & 0\\
0&1\end{pmatrix},
\end{gathered}
\end{equation}
 we obtain its eigenvalues
\begin{equation}
\label{m12}
m_{1,2} = \frac{1}{2}{m_R}\left( {\sqrt {1 + 4\frac{{m_D^2}}{{m_R^2}}} \mp 1} \right).
\end{equation}

The matrix $P$ is introduced in (\ref{diag}) in order to change the sign of the mass $m_1$ so that both masses $m_1$ and $m_2$ are positive.
The corresponding eigenvectors of the mass matrix (Majorana fields with definite masses) are written as follows:
\begin{equation}
\label{ev}
\begin{gathered}
\begin{pmatrix}\nu_1\\ \nu_2\end{pmatrix} = U^+\begin{pmatrix}\chi\\N\end{pmatrix},\quad U^+ = \begin{pmatrix} i\cos\theta& - i\sin\theta\\
\sin\theta&\cos\theta \end{pmatrix},\\
\nu _{1} = i\chi \cos \theta  - iN\sin \theta = - \nu_1^c ,\\
\nu _{2} =  \chi \sin \theta  + N\cos \theta = \nu_2^c \, ,  \\
\tan(2\theta)  = \frac{2m_{D}}{m_{R}}.
\end{gathered}
\end{equation}
As a result, (\ref{DM}) takes the form of the standard Majorana
mass term (cf. (\ref{Maj}))
\begin{equation}
\label{MMS}
{\cal L}_{DM} = - \frac{1}{2}\sum\limits_{k = 1}^{2}m_{k}{\bar \nu}_{k}\nu_{k}.
\end{equation}

The initial flavor fields (included in the Lagrangian (\ref{NR})
and mass term (\ref{DM})) turn out to be superpositions of Majorana fields with certain masses:
\begin{equation}
\label{FM}
\begin{gathered}
\begin{pmatrix}\nu_L\\N^c_R\end{pmatrix} = U\begin{pmatrix}\nu_{1L}\\ \nu_{2L}\end{pmatrix},\\
\nu_L = - i\nu_{1L}\cos\theta + \nu_{2L}\sin\theta,\\
N_{R}^{c} =  i\nu_{1L}\sin\theta + \nu_{2L}\cos\theta.
\end{gathered}
\end{equation}

In the case of the heavy right-handed neutrino $N_R$, we have
$m_R \gg m_D$, and from (\ref{m12}) and (\ref{ev}) we obtain
\begin{equation}
\label{seesaw}
{m_1} \simeq \frac{{m_D^2}}{{{m_R}}} \ll {m_D} \, ,
\quad \quad {m_2} \simeq {m_R} \, ,
\quad \quad \theta  \simeq \frac{{{m_D}}}{{{m_R}}} \ll 1.
\end{equation}
Therefore, if $m_D$ is of the order of the mass of a charged fermion, then the mass of a light neutrino turns out to be very small even for the Yukawa coupling constant $y_{\nu} \sim 1$ due to the
interaction with the heavy Majorana neutrino.
This is the essence of the seesaw mechanism. In this case, as can be seen from (\ref{FM}) and (\ref{seesaw}), the active flavor neutrino $\nu_L$ contains a small admixture of the heavy Majorana neutrino.
For example, for $m_D = m_t \simeq 173~\text{GeV}$ and $m_1 = 0.05~\text{eV}$ we find ${m_R} \simeq m_t^2/{m_1} \simeq 6 \times {10 ^{14}}~\text{GeV}$ (see (\ref{hy}) and (\ref{msc})), which is a typical grand unified scale \cite{em,nag,lang}.

In Appendix B, we consider a generalization of SSM
corresponding to the modification of the mass matrix in (\ref{MM}):
\[
\begin{pmatrix} 0&m_D\\m_D&m_R\end{pmatrix} \rightarrow \begin{pmatrix} m_L&m_D\\m_D&m_R\end{pmatrix} \; ,
\]
where $m_L$ is small but is not equal to zero.

\label{sec:Leff}
\section{Effective Lagrangians for Seesaw Mechanism}

{\bf 2.1.} We consider the effective Lagrangian ${\cal L}_{\text {eff}}$ corresponding to the model (\ref{NR}) and suitable for energies
$E \ll m_{R}$. It is determined by functional integration over heavy Majorana fields:
\begin{equation}
\label{Le}
e^{iS_{\text{eff}}} = \exp \left( {i\int {{d^4}x{{\cal L}_{\text{eff}}}(x)} } \right) = \int [dN] [d\bar N]\exp \left(i\int {{d^4}x{{\cal L}_{\nu N}}(x)}\right),
\end{equation}
where the Lagrangian (\ref{NR}) is conveniently
represented in the form
\begin{equation}
\label{LNJ}
\begin{gathered}
{\cal L}_{\nu N} =\bar N{\hat K}N - \bar NJ - {\bar J}N,\\
\hat K := \frac{1}{2}( i\gamma ^{\mu }\partial _{\mu } - m_{R}),\\
J :=  \frac{1}{2}y_{\nu }\left({\tilde \varphi }^ {+ }L _{L} + {\tilde \varphi }^{T}L _{L}^{c} \right),\quad {\bar J} :=
\frac{1}{2}y_{\nu }\left({\bar L }_{L}{\tilde \varphi}  + {\bar L} _{L}^{c}{\tilde \varphi}^{*}\right).
\end{gathered}
\end{equation}
For deriving (\ref{LNJ}) we use the relations
\begin{equation}
\label{rel2}
\begin{gathered}
N = N^{c} \, ,\quad \quad {L}_{L}^{c} = C{\bar L}_{L}^{T} \, ,
\quad \quad {\bar L}_{L}^{c} = - {L}^{T}C^{- 1} \, ,   \\
{\bar L }_{L}{\tilde \varphi}N = -N^{T}{\tilde \varphi}^{T}
C^{- 1}C{\bar L }_{L}^{T}={\bar N}{\tilde \varphi}^{T}{L}_{L}^{c} \, ,
\quad \;\;
{\bar N}{\tilde \varphi}^{+}{L}_{L} = {\bar L}_{L}^{c}{\tilde \varphi}^{*}N.
\end{gathered}
\end{equation}
The integration in (\ref{Le}) over fermion fields
(taking into account methods of \cite{pop}) gives,
in view of (\ref{LNJ}),
\begin{equation}
\label{Seff}
e^{iS_{\text{eff}}} = \text{det}{\hat K}\cdot \exp
\left( - i\int{d^{4}x{\bar J}{\hat K}^{- 1}J}\right).
\end{equation}
The determinant $\text{det}{\hat K}$ in (\ref{Seff}) does not depend on the fields, and taking it into account
adds only a constant term to the effective action, which can be omitted. As a result, we obtain the effective Lagrangian in the form
\begin{equation}
\label{Leff}
{\cal L}_{\text{eff}} = - {\bar J}{\hat K}^{- 1}J.
\end{equation}
From here, putting ${\hat K}^{- 1} \simeq - 2/m_R$ (in the leading order of expansion over $1/m_R$) and taking into account (\ref{LNJ}), we obtain
\begin{equation}
\label{LO5}
{\cal L}_{\text{eff}} =  \frac{y_{\nu }^{2}}{2m_R}\left( {{{\bar L }_L}\tilde \varphi {{\tilde \varphi }^T}L _L^c + \bar L _L^c{{\tilde \varphi }^*}{{\tilde \varphi }^ + }{L _L}} \right),
\end{equation}
 which is a special case of the ${\cal O}^{(5)}$ operator ({\ref{O5}).
 After substitution of (\ref{mD}), the effective Lagrangian
 (\ref{LO5}) gives
 the corresponding mass term for the Majorana field $\chi$ introduced in (\ref{rel}):
\begin{equation}
\label{LM}
\begin{gathered}
{\cal L}_M = \frac{{y_\nu ^2{v^2}}}{{4{m_R}}}\left( {{{\bar \nu }_L}\nu _L^c + \bar \nu _L^c{\nu _L}} \right) = \frac{1}{2}m{\bar \chi} \chi ,\\
m = \frac{{m_D^2}}{{{m_R}}},\quad \chi  = {\nu _L} + \nu _L^c =  \chi^{c} .
\end{gathered}
\end{equation}
The ``incorrect'' sign of the mass term (\ref{LM}) is removed by
the transition to the Majorana field of negative charge parity $\chi'$:
\begin{equation}
\label{ii}
\chi  \to \chi' = i\chi  = - \chi'^{c}.
\end{equation}
Indeed, taking into account the relations
(see  Eq. (\ref{bpc}) in Appendix A)
\begin{equation}
\label{rs}
{\bar\chi}\chi ={\bar\chi}^c\chi = \chi^{T}C\chi = (-i\chi')^{T}C(-i\chi') = - {\bar{\chi'}}\chi',
\end{equation}
we obtain
\begin{equation}
\label{LM1}
{\cal L}_M = -\frac{1}{2} m {\bar{\chi'}}\chi'.
\end{equation}

Formulas (\ref{LM}) and (\ref{LM1}) agree with (\ref{ev}) -- (\ref{seesaw}) in the leading order of
the expansion of (\ref{Leff}) over
$1/m_R$, as it should be.

\vspace{0.1cm}

{\bf 2.2.} Now we consider an extension of the SM with three lepton generations interacting with three heavy right-handed neutrinos $N_{jR}~(j = 1,2,3)$. The corresponding interaction Lagrangian
which generalizes (\ref{NR}) has the form \cite{xi,bro} (we use
here and below the concise matrix notation):
\begin{equation}
\label{SMNR}
\begin{gathered}
{\cal L}_{\ell N} = {\bar N}_Ri\gamma^{\mu} {\partial}_{\mu}N_R - \frac{1}{2}\left( {\bar N_R^c{M_R}{N_R} + {{\bar N}_R}{M_R^*}N_R^c} \right) \\
-  {{{\bar L }_L}\tilde \varphi {Y^\nu}{N_R} - {{\bar N}_R}{Y^{\nu +}}{{\tilde \varphi }^ + }{L_L}},
\end{gathered}
\end{equation}
where $N_R = (N_{jR})$, the matrix of Yukawa coupling constants
is denoted as $Y^{\nu} = (Y_{\alpha j}^{\nu})$ and
$M_R = (M_{Rjk})$ is the complex symmetric
$3\times 3$ mass matrix: $M_{Rjk} = M_{Rkj}$ ($j,k = 1,2,3$).

The matrix $M_R$ is diagonalized by using the unitary matrix $U$:
\begin{equation}
\label{M}
U^{T}M_{R}U = M = \text{diag}(M_1,M_2,M_3),\quad M_R = U^*MU^+,
\end{equation}
where $M_j$ are real {\it nonnegative} numbers. This statement is a special case of Takagi's diagonalization theorem (see \cite{drei}, Appendix D, and references therein). This theorem
 states that for any complex symmetric $n\times n$-matrix $M_S$ there exists a unitary $n\times n$-matrix $U$ such that
\[
U^{T}M_{S}U = M = {\rm diag}(m_1,m_2, ... ,m_n),
\]
where all $m_j$ are real and nonnegative. We prove this
important theorem in Appendix C by modifying the proof from
\cite{drei}.

Applying (\ref{M}), we represent the Lagrangian (\ref{SMNR}) in the mass basis of heavy Majorana neutrinos:
\begin{equation}
\label{SMNR1}
{\cal L}_{\ell N} = \frac{1}{2}{\bar N}_j\left( {i\gamma  \cdot \partial  - {M_j}} \right){N_j} - {{{\bar L }_{\alpha L}}\tilde Y_{\alpha j}^\nu \tilde \varphi {N_j} - {{\bar N}_j}\tilde Y_{j\alpha }^{\nu *}{{\tilde \varphi }^ + }{L _{\alpha L}}}.
\end{equation}
Here
\begin{equation}
\label{tilde}
{N_j} = {{\tilde N}_{jR}} + \tilde N_{jR}^c = N_j^c,\quad {{\tilde N}_R} = ({{\tilde N}_{jR}}) = {U^+}{N_R},\quad {{\tilde Y}^\nu } = {Y^\nu }{U}.
\end{equation}
Using obvious generalizations of relations (\ref{rel2}),
we represent (\ref{SMNR1}) in a form analogous to (\ref{LNJ}):
\begin{equation}
\label{SMNR2}
\begin{gathered}
{\cal L}_{\ell N} = {\cal L}_{\nu N} =\bar N{\hat K}N - \bar NJ - {\bar J}N,\\
\hat K = \frac{1}{2}( i\gamma\cdot \partial  - M) = ({\hat K}_{jk}),\quad {\hat K}_{jk} = \frac{1}{2}( i\gamma\cdot \partial  - M_j)\delta_{jk},\\
J =  \frac{1}{2}\left({\tilde Y}^{\nu +}{\tilde \varphi }^ {+ }L _{L} + {\tilde Y}^{\nu T}{\tilde \varphi }^{T}L _{L}^{c} \right),\quad
{\bar J} = \frac{1}{2}\left({\bar L }_{L}{\tilde \varphi}{\tilde Y}^{\nu}  + {\bar L} _{L}^{c}{\tilde \varphi}^{*}{\tilde Y}^{\nu *}\right).
\end{gathered}
\end{equation}

Now we substitute (\ref{SMNR2}) into (\ref{Le}) and use
(\ref{Seff}), (\ref{Leff}) to obtain an effective Lagrangian
which generalizes (\ref{LO5}):
\begin{equation}
\label{Leff1}
{\cal L}_{\text{eff}} = \frac{1}{2}{{\bar L}_L}\tilde \varphi {{\tilde Y}^\nu }{M^{ - 1}}{{\tilde Y}^{\nu T}}{{\tilde \varphi }^T}L _L^c + \text{H.c.}
\end{equation}
Then, applying relations (\ref{M}) and (\ref{tilde}), we transform (\ref{Leff1}) to the form of the ${\cal O}^{(5)}$
Weinberg operator (\ref{O5}):
\begin{equation}
\label{Leff2}
\begin{gathered}
{\cal L}_{\text{eff}} = z_{\alpha\beta}{{\bar L}_{\alpha L}}\tilde \varphi {{\tilde \varphi }^T}L _{\beta L}^c + \text{H.c.},\\
z_{\alpha\beta} = \frac{1}{2}{\left( {{Y^\nu }M_R^{ - 1}{Y^{\nu T}}} \right)_{\alpha \beta }}.
\end{gathered}
\end{equation}

After spontaneous symmetry breaking, the interaction with the Higgs doublet (\ref{Leff2}) generates the Majorana mass matrix of light neutrinos (cf. the case of one lepton generation (\ref{LM})):
\begin{equation}
\label{m}
m_{\alpha\beta} = - z_{\alpha\beta}v^2 = - ({M_D}M_R^{-1}M_D^T)_{\alpha\beta}, \quad M_D = Y^{\nu}\frac{v}{\sqrt{2}},
\end{equation}
so that the mass term in the Lagrangian has the form
\begin{equation}
\label{LM2}
{\cal L}_M = - \frac{1}{2}\left(m_{\alpha\beta}
{\bar \nu _{\alpha L}}\nu _{\beta L}^c + \text{H.c.}\right).
\end{equation}
Diagonalization of the matrix (\ref{m}) leads
the mass term (\ref{LM2})
to the standard form corresponding to three light
Majorana neutrinos (see (\ref{Maj})).

\vspace{0.1cm}

{\bf 2.3.} The seesaw mechanism discussed above
generates the Majorana neutrino mass. However, the nature
of the neutrino mass (whether it is Majorana or Dirac) has not yet been established experimentally\cite{pdg}. Therefore, it is interesting to consider this mechanism for Dirac neutrinos (see \cite{chen} and the literature cited there).

For a simple model with one lepton generation, the corresponding Yukawa part of the full Lagrangian has the form (cf. (\ref{NR})):
\begin{equation}
\label{nD}
\begin{gathered}
{\cal L}_{\nu D} =   - y_{\nu }{\bar L}_L{\tilde \varphi }N_{R} - m_{R}{\bar N}_{L}\nu _{R}  - m_{N}{\bar N}_{L}N_{R} + \text{H.c.} \\
= - y_{\nu }\left({\bar L}_L{\tilde \varphi }N +{\bar N}{\tilde \varphi }^{+}L_{L}\right) -  m_{R}\left({\bar N}\nu _{R} +{\bar \nu }_{R} N\right) - m_{N}{\bar N}N,
\end{gathered}
\end{equation}
where the Dirac bispinor $N = N_L + N_R$ is a singlet
with respect to the SM gauge group and describes ``heavy'' degrees of freedom under the assumption that the mass $m_N$ is significantly larger than $m_R$ and the typical electroweak scale $v$ (see (\ref{vac})).

The effective Lagrangian is obtained by substituting into (\ref{nD}) expressions for ''heavy'' fields via ``light'' fields (that is equivalent, in the accepted leading order of expansion over $1/m_N$, to
the integration over heavy fermion fields in the
generating functional, see \cite{bro,alt}):
\[
N =  - \frac{1}{{{m_N}}}\left( {{y_\nu }{{\tilde \varphi }^ + }{L_L} + {m_R}{\nu _R}} \right),\quad \bar N =  - \frac{1}{{{m_N}}}\left( {{y_\nu }{{\bar L}_L}\tilde \varphi  + {m_R}{{\bar \nu }_R}} \right) \, .
\]
These expressions follow from the equations of motion in the static approximation (neglecting the contribution of kinetic terms in the Lagrangian, which is justified in the energy region $E \ll m_N$):
\[
\frac{\partial {\cal L}_{\nu D}}{\partial \bar N} = 0,\quad \frac{\partial {\cal L}_{\nu D}}{\partial N} = 0.
\]
As a result, we obtain (cf. (\ref{LO5}))
\begin{equation}
\label{LeffD}
{\cal L}_{\text{eff}} = \frac{{{y_\nu }{m_R}}}{{{m_N}}}\left( {{{\bar L}_L}\tilde \varphi {\nu _R} + {{\bar \nu }_R}{{\tilde \varphi }^ + }{L_L}} \right).
\end{equation}
After spontaneous symmetry breaking (see (\ref{vac})),
from (\ref{LeffD}) the Dirac mass term follows (cf. (\ref{LM})--(\ref{LM1}))
\begin{equation}
\label{1D}
\begin{gathered}
{\cal L}_{D} = {m_\nu }\left( {{{\bar \nu }_L}{\nu _R} + {{\bar \nu }_R}{\nu _L}} \right) = {m_\nu }{\bar \nu}\nu  =  - {m_\nu }\bar \nu '\nu ',\\
\nu ' = {\gamma ^5}\nu = {\gamma ^5}\left( {{\nu _L} + {\nu _R}} \right) = \nu _R - \nu _L,\\
m_{\nu} = \frac{m_{L}m_{R}}{m_{N}},\quad {m_L} = {y_\nu }\frac{v}{{\sqrt 2 }},\\
{m_\nu } \ll {m_{L,R}} \ll {m_N}.
\end{gathered}
\end{equation}
Here the correct sign of the mass is provided by
making use of the $\gamma^5$-transformation\footnote{We stress
that $\gamma^5$-transformation of the Majorana fields
does not change the sign of the Majorana mass term. That is why,
in the Majorana neutrino case (see item {\bf 2.2} above),
we apply another transformation (\ref{ii}) to change
the sign
of the Majorana mass term (\ref{rs}) and
provide the positive mass for Majorana neutrinos.}
$\nu \to \gamma^5 \, \nu$ of the Dirac bispinor
\cite{chli}:
\[
\bar \nu \nu  = {\nu ^ + }{\gamma ^0}\nu  =
\nu {'^ + }{\gamma ^5}{\gamma ^0}{\gamma ^5}\nu ' =  - \bar \nu '{\left( {{\gamma ^5}} \right)^2}\nu ' =  - \bar \nu '\nu '.
\]

There are three possible relationships for the mass parameters:
\[
1)~ m_L \sim m_R, \quad 2) ~m_L \ll m_R, \quad 3)~ m_L \gg m_R.
\]
As it was shown in \cite{chen}, the case 3), called the undemocratic Dirac seesaw mechanism (with an appropriate generalization to
several lepton generations), can be used to describe baryon asymmetry of the Universe, as well as the stability of the dark matter.

{\bf 2.4.} Following the work \cite{chu}, we consider a generalization of the model (\ref{nD}) to three lepton generations (for another generalization, see \cite{chen}).
This generalization is based on
{the extension
of the  gauge group
${\rm SU(3)}_c\times {\rm SU(2)}_L\times
{\rm U(1)}_Y$ by the additional symmetry
  group ${\rm Z}_4\times {\rm Z}_2$. Besides the standard fermions involved in the SSM (left doublets $L_{\alpha L}$,
  right singlets $\ell_{\alpha R}$ and
  right singlets $\nu_{\alpha R}$), the model includes three heavy Dirac fermions $N_i$ and three scalars (in addition to the Higgs doublet $\varphi$)
  $\chi, \zeta, \eta$, which are gauge singlets.
The action of the discrete
group ${\rm Z}_4 \times {\rm Z}_2$
on the fields of the model is the following:
$$
 \begin{array}{c}
\bar{L}_{\alpha L}  \to
({\bf z}^3 \times {\bf 1}) \, \bar{L}_{\alpha L} \, , \;\;\;
\ell_{\alpha R} \to ({\bf z} \times {\bf 1}) \, \ell_{\alpha R}
\, , \;\;\; \nu_{\alpha R} \to ({\bf z} \times (-{\bf 1})) \, \nu_{\alpha R}
\, , \\ [0.2cm]
\bar{N}_{i L}  \to
({\bf z}^3 \times {\bf 1}) \, \bar{N}_{i L} \, , \;\;\;
N_{i R} \to ({\bf z} \times {\bf 1}) \, N_{i R} \, , \;\;\;
\varphi \to ({\bf 1} \times {\bf 1})
\, \varphi \, , \\ [0.2cm]
\chi \to ({\bf 1} \times (-{\bf 1})) \, \chi  \, , \;\;\;
\zeta \to ({\bf z} \times {\bf 1}) \, \zeta  \, , \;\;\;
\eta \to ({\bf z}^2 \times {\bf 1}) \, \eta  \, ,
\end{array}
$$
where ${\bf z} = e^{i \frac{\pi}{2}}$ is the
primitive fourth root of unity ${\bf z}^4={\bf 1}$.
 Thus,} the scalar $\chi$ is uncharged with respect to
 the group ${\rm Z}_4$ but is odd with respect to ${\rm Z}_2$,
the other two scalars are
{un}charged in ${\rm Z}_2$.

  The  cyclic group ${\rm Z}_4$
(which describes a discrete analogue of the lepton number) prohibits the Majorana terms in the Lagrangian and provides the stability of
candidates for dark matter particles, while the group
${\rm Z}_2$
provides a seesaw mechanism for the generation of Dirac neutrino mass,
prohibiting tree amplitudes that connect left-handed and right-handed
 neutri\-nos (for discrete groups ${\rm Z}_n$ see \cite{ir}).

The part of the Lagrangian of the model responsible for the generation of the mass of light (active) neutrinos
 {and invariant under the action of the group
 $Z_4 \times Z_2$}
has the form
\begin{equation}
\label{3D}
{\cal L}_{\nu N\chi } =  - {f_{\alpha i}}{\bar L_{\alpha L}}\tilde \varphi {N_{iR}} - {g_{i\alpha }}{\bar N_{iL}}\chi {\nu _{\alpha R}} - M_{ij}{\bar N_{iL}}{N_{jR}} + \text{H.c.},
\end{equation}
where $f_{\alpha i},\, g_{\alpha i}$ are constants,
$M_{ij}$ is the mass matrix,
$\alpha = e, \mu, \tau$ and $i = 1, 2, 3$.
The corresponding effective Lagrangian is obtained by means of
 obvious generalization of the method outlined in Section 2.3
 (we use the index free matrix notation):
\begin{equation}
\label{Leff3D}
{\cal L}_{\text{eff}} = {{\bar L}_L} \bigl( f\tilde \varphi
\; {M^{ - 1}} \; g\chi \bigr) {\nu _R} +  \text{H.c.} \;
\end{equation}
After spontaneous symmetry breaking, the scalars $\tilde \varphi$ and $\chi$ obtain vacuum expectation values (see (\ref{vac})):
\begin{equation}
\label{vac2}
\langle 0|\tilde\varphi|0\rangle = (v/\sqrt{2},0)^{T}, \quad \langle 0|\chi|0\rangle = u,
\end{equation}
and as a result, the Dirac mass term is generated
\[
{\cal L}_{D} = {{\bar \nu }_L}{M_\nu }{\nu _R} + \text{H.c.},
\]
where the mass matrix is (cf. (\ref{1D}))
\begin{equation}
\label{m3D}
{M_\nu } = \frac{vu}{{\sqrt 2 }}fM^{ - 1}g \; .
\end{equation}

Note that the ${\rm Z}_4$-charged scalars $\zeta$ and $\eta$
(in contrast to the neutral $\varphi$ and $\chi$) have zero vacuum
expectation values, , so that the group ${\rm Z}_4$ remains
unbroken after spontaneous breaking of electroweak symmetry, while
the vacuum expectation value of the scalar $\chi$ (see (\ref{vac2})),
which is odd with respect to ${\rm Z}_2$, breaks ${\rm Z}_2$-symmetry spontaneously, which generates small Dirac masses (see (\ref{m3D}) for $|M_{ij}|\gg v, u$).

Besides, the analysis shows \cite{chu} that $\zeta$ particles turn out to be stable, and therefore can be considered as candidates
for dark matter particles, while $\eta$ particles are unstable.

\label{sec:con}
\section{Conclusion}

We have considered the Majorana and Dirac versions of the seesaw
 mechanism (of type I) for generating the mass of active light neutrinos, which is based on the extension of the SM by adding heavy neutrinos that have a Yukawa interaction with standard flavor neutrinos.
By integrating over the
''heavy'' fields in the generating functional of the theory, the corresponding low-energy effective Lagrangians were obtained, which, after spontaneous breaking of electroweak symmetry, lead
to the mass terms of light neutrinos.

We stress the fundamental difference between the mechanisms of mass generation of charged leptons (and quarks) and light neutrinos: the masses of the former are determined by the products of the electroweak scale $v$ and the corresponding {\it dimensionless} Yukawa coupling constants (see (\ref{mk})), while the smallness of the masses of
the second (active) neutrinos is ensured by the introduction of the
 {\it dimensional} parameter --- the large mass scale of new physics $\Lambda$, so that the mass value
 is suppressed by the small ratio $v/\Lambda$ (see (\ref{msc})).

In the case of the considered seesaw mechanism (SSM), the scale $\Lambda$ represents the scale of the masses $M$ of heavy neutrinos introduced during the extension of the Standard Model.
However, a natural question arises (see, for example, \cite{moh2}) about the generation of the scale $M$ itself, which was
introduced above ''by hand''
(see $m_R$ in (\ref{NR}) and $m_N$ in (\ref{nD})).
As it can be seen from (\ref{msc}), the scale of $\Lambda$ coincides in order of magnitude with the typical scale of Grand Unified Theories (GUT). An example of an extension of the SM leading to SSM is GUT based on the gauge group ${\rm SO}(10)$ \cite{g-m}.
There are many ways of spontaneous breaking of this group up to the SM group ${\rm G}_{\rm SM}={\rm  SU}(3)_c  \times
{\rm SU}(2)_L  \times {\rm U}(1)_Y$ with its subsequent breaking on the scale $v$. Besides, the minimal way contains two steps (see \cite{dep}, where the supersymmetric ${\rm SO}(10)$ theory is considered):
\[
{\rm SO}(10)\xrightarrow{\Lambda _{\rm GUT}} {\rm G}_{\rm SM}\xrightarrow{v}{ {\rm SU}\left( 3 \right)_c}  \times { {\rm U}(1)_{\rm em}}.
\]

Non-supersymmetric ${\rm SO}(10)$-GUT is considered in the work \cite{fu}, where, in particular, it is shown how, in the case of spontaneous breaking of ${\rm SO}(10)$, the SSM of type I arises for
light neutrinos.

Left-right symmetric theories also lead to
the SSM \cite{moh,moh2}.

Note that recently a new type of the seesaw mechanism \cite{shr}, based on models with extra dimensions (see, e.g., \cite{nag} and references therein), has been proposed. In these models, the 4-dimensional Minkowski space is embedded in a space of dimension $4 + n$, with $n$ extra spatial dimensions compactified on a length scale $L$, which corresponds to the energy scale $\Lambda_L = 1/L $. Integration over the extra dimensions leads to the Lagrangian of the low-energy effective field theory, which after spontaneous symmetry breaking gives the Majorana mass term for active neutrinos with small masses of order $m_{\nu} \sim m_{D}^2/\Lambda_L \sim 10^{- 3}~\text{eV}$ for $m_D = m_e$  and $\Lambda_L = 100~\text{TeV}$ (cf. (\ref{seesaw})).

\vspace{0.1cm}

The Majorana SSM leads to non-conservation of the lepton number $L$, changing it by two units (see (\ref{Leff1}) and (\ref{LM2})).
 This opens up the possibility of observing numerous
 physical processes with $|\Delta L| = 2$ induced by Majorana neutrinos: neutrinoless double beta decay of nuclei (see review \cite{cir}) and its analogs --- semileptonic decays of mesons with the birth of a pair of leptons with identical electric charges (dileptons) \cite{ali1,ali2 }, production of dileptons in deep inelastic proton-proton and lepton-proton collisions at high-energy colliders (see, for example, \cite{ali1,ali3}), etc. The search for such processes is one of the important areas of researches in particle physics \cite{pdg}.

 Heavy Majorana neutrinos can play a significant role in cosmology: their decays with CP violation at the early stages of the Universe evolution lead to leptonic asymmetry, that, due to a special non-perturbative electro\-weak interaction of leptons and quarks with non-conservation of lepton and baryon numbers, is transformed into baryon asymmetry. This mechanism for genera\-ting
  baryon asymmetry in the Universe is called
  the leptogenesis (see \cite{goru} and the references therein).
 An extension of the Standard Model by adding three heavy right-handed neutrinos is called the neutrino minimal Standard Model ($\nu$MSM).
 Its applications in cosmology, including the prob\-lems of baryogenesis and dark matter, are considered, for example, in the works \cite{sh1,sh2}.

 In what concerns the Dirac SSM, as it was indicated above,
 by considering
 the corresponding extension of the SM, new scalar singlets are introduced, and one of these singlets
  can serve as a candidate for the role of
  particles of the stable dark matter.

Thus, the small masses of active neutrinos serve as
a clear signal of new physics, which is not described by the Standard Model \cite{moh2}.

\label{sec:app}
\section*{Appendix A. Weyl, Dirac and Majorana spinors
in Minkowski space $\mathbb{R}^{1,3}$}

\renewcommand{\theequation}{A.\arabic{equation}}
\setcounter{equation}{0}

To describe fermion fields with spin $1/2$, two-component
complex Weyl spinors are used, which are transformed independently
in the fundamental (spinor {$\xi_{a}$, $a=1,2$})
and antifundamental, or complex conjugate to the fundamental, (spinor
{$\eta^{\dot a}$, ${\dot a}=1,2$}) representations of the group $\rm SL(2,\mathbb{C})$ (here we follow \cite{drei,Novo,ir2}):
\begin{equation}
\label{weisp}
\begin{gathered}
  {{\xi '}_a} = {A_a}\,^b{\xi _b}\, ,\quad {{\eta '}^{\dot a}} =
  {\tilde A}^{\dot a}\,_{\dot b}{\eta ^{\dot b}} \, ; \\
  \tilde A = {\left( {{A^{ - 1}}} \right)^ + },\quad \det A = 1.
\end{gathered}
\end{equation}
As it is known, $\rm SL(2,\mathbb{C})$ is a double covering group for the proper orthochronous Lorentz group {$\rm SO^{\uparrow}(1,3)$}.
 The complex ${(2\times 2)}$-matrix
$A \in \rm SL(2,\mathbb{C})$ corresponds to a real pseudo-orthogonal
 $(4\times4)$-matrix $\Lambda \in \rm SO^{\uparrow}(1,3)$:
\begin{equation}
\label{CL}
{\Lambda _{\mu \nu }} = {g_{\mu \lambda }}{\Lambda ^\lambda }_\nu  = \frac{1}{2}{\text{tr}}\left( {{{\sigma }_\mu }A{\tilde\sigma _\nu }{A^ + }} \right) \, , \;\;\;\; {\mu,\nu,... = 0,1,2,3} \; ,
\end{equation}
where {$g_{\mu \lambda } = {\rm diag}(+1,-1,-1,-1)$} and two sets of four $2\times 2$-matrices were introduced, including Pauli matrices
 $\boldsymbol{\sigma} = (\sigma_1,\sigma_2,\sigma_3)$ and unit
  matrix ${\sigma _0} = I$:
{\begin{equation}
\label{sigma}
{{\tilde \sigma }^\mu } =  ||({\tilde \sigma }^\mu)^{{\dot c} a}|| = \left(I, - \boldsymbol{\sigma } \right),\quad \quad
  \sigma ^\mu =
||(\sigma ^\mu)_{a{\dot c}}|| = \left(I,\boldsymbol{\sigma}\right),
\end{equation}
{\begin{equation}
\label{sigma1}
\sigma_1 = \begin{pmatrix}
  0 & 1 \\
  1 & 0
\end{pmatrix} \; , \;\;\;\;
\sigma_2 = \begin{pmatrix}
  0 & -i \\
  i & 0
\end{pmatrix} \; , \;\;\;\;
\sigma_3 = \begin{pmatrix}
  1 & 0 \\
  0 & -1
\end{pmatrix} \; .
\end{equation}}
Note that the double covering means that two elements of the group $\rm SL(2,\mathbb{C})$ correspond, as follows from (\ref{CL}), to one element of the Lorentz group $\rm SO^{\uparrow }(1,3)$: $\pm A \to \Lambda$.
That is why, the spinor representations of the Lorentz group are called double-valued and the physical observables may not be the spinor fermion
fields themselves but their bilinear combinations.

The Dirac 4-component spinor (bispinor) $\psi$ is composed of two Weyl spinors (\ref{weisp}):
\begin{equation}
\label{dirspin}
\psi  =
 \begin {pmatrix}
 \xi_{_{(L)}}  \\
  \eta_{_{(R)}}
\end{pmatrix} =
\begin {pmatrix}
  \xi_a  \\
  \eta^{\dot a}
\end{pmatrix}
\, ,\quad  \xi_{_{(L)}}  := (\xi_a) = \begin{pmatrix}
  \xi_1\\
  \xi_2
\end{pmatrix}\, , \quad \eta_{_{(R)}}  :=
(\eta ^{\dot a }) = \begin{pmatrix}
  \eta ^{\dot 1} \\
  \eta ^{\dot 2}
\end{pmatrix},
\end{equation}
which corresponds to the Weyl representation of gamma matrices (see (\ref{sigma})):
\begin{equation}
\label{gwei}
\begin{gathered}
\gamma ^\mu = \begin{pmatrix}
  0&\sigma ^\mu \\
  {\tilde\sigma }^\mu &0
\end{pmatrix}, \quad \sigma^\mu  = (I, \boldsymbol{\sigma}),
\quad \tilde{\sigma}^\mu = (I, - \boldsymbol{\sigma}),\\
\gamma ^0 =  \begin{pmatrix}
  0&I \\
  I&0
\end{pmatrix}, \quad \boldsymbol{\gamma}= (\gamma ^k) = \begin{pmatrix}
  0&  \boldsymbol{\sigma} \\
  - \boldsymbol{\sigma}&0
\end{pmatrix}, \quad
\gamma^5 = i\gamma^0\gamma^1\gamma^2\gamma^3 =
\begin{pmatrix} - I&0\\0& I\end{pmatrix}.
\end{gathered}
\end{equation}
The Dirac conjugated bispinor
to the bispinor (\ref{dirspin}) has the
form\footnote{{The Dirac conjugated bispinor
$\bar \psi$ is constructed in a such way
 that it can be covariantly contracted with the spinor
 $\psi$. In the definition of $\bar \psi$,
the matrix $\gamma^0$ only formally coincides with the Dirac matrix $\gamma^0$, as it can be seen from the arrangement of its
dotted and undotted spinor indices; see (\ref{sigma}).}}
\begin{equation}
\label{cdir}
\begin{gathered}
  \bar \psi  = {\psi ^ + }{\gamma ^0} = \left( {{{\bar \xi }_{\dot a}},
  {{\bar \eta }^a}} \right){\gamma ^0} = \left( {{{\bar \eta }^a},{{\bar \xi }_{\dot a}}} \right) \, ,
  \end{gathered}
\end{equation}
\begin{equation}
\label{cdir2}
  {{\bar \xi }_{\dot a}} = {\left( {{\xi _a}} \right)^*} \, , \;\;\;\;
  {{\bar \eta }^a} = {\left( {{\eta ^{\dot a}}} \right)^*}.
\end{equation}

The left-handed $\psi_L$ and the right-handed $\psi_R$ bispinors,
which compose the bispinor (\ref{dirspin}), are expressed
in terms of Weyl spinors as follows:
\begin{equation}
\label{LR}
\begin{gathered}
\psi_L = {
\frac{1}{2}(1 - \gamma^5)} \psi =
\begin{pmatrix} \xi_a\\0\end{pmatrix} \, , \quad
\psi_R = {\frac{1}{2}(1 +
 \gamma^5)} \psi =  \begin{pmatrix} 0\\ \eta ^{\dot a}\end{pmatrix}, \\
\psi = \psi_L + \psi_R,
\end{gathered}
\end{equation}
where the matrix $\gamma^5$ is defined in (\ref{gwei}).

The covering group $\rm SL(2,\mathbb{C})$ of the
 Lorentz group $\rm SO^{\uparrow}(1,3)$
 acts on the bispinor (\ref{dirspin}) according to (\ref{weisp}):
\begin{equation}
\label{SLor}
\begin{gathered}
\psi(x) \to \psi^{\prime}(x^{\prime}) = \begin{pmatrix} A&0\\0&{\tilde A}\end{pmatrix}\!\psi(x)  \equiv D(A)\psi(x), \\
D(A) = \exp \left(\frac{i}{4}\omega_{\mu\nu}\Sigma^{\mu\nu}\right), \quad \Sigma^{\mu\nu} = \frac{i}{2}
[\gamma^{\mu}, \, \gamma^{\nu}] = { \frac{i}{2} \begin{pmatrix}\sigma^{\mu\nu}&0\\
 0&\tilde{\sigma}^{\mu\nu} \end{pmatrix} },\\
(\sigma^{\mu\nu})_a^{\;\; b} =
(\sigma^{\mu})_{a \dot{c}}
(\tilde{\sigma}^{\nu})^{\dot{c}b} -
(\mu \leftrightarrow \nu), \;\;\;
(\tilde{\sigma}^{\mu\nu})_{\;\; \dot b}^{\dot a} =
(\tilde{\sigma}^{\nu})^{\dot{a}c}
(\sigma^{\mu})_{c\dot{b}} -
(\mu \leftrightarrow \nu),
\end{gathered}
\end{equation}
where the matrix $\omega_{\mu\nu} = - \omega_{\nu\mu}$
is composed of 6 real parameters of the Lorentz group.
 This follows from the fact that any
complex ${(2\times 2)}$-matrix $A\in \rm SL(2,\mathbb{C})$
is represented as
 a product of Hermitian and unitary matrices
 which respect\-ively correspond to a real
 pseudo-orthogonal $(4\times4)$ boost matrix
 $\Lambda \in \rm SO^{\uparrow}(1,3)$
 and a pure three-dimensional rotation matrix in
 $\mathbb{R}^{3} \subset \mathbb{R}^{1,3}$.
 Taking into account the Lorentz transformation
 for the coordinates ${x'^\mu } = {\Lambda ^\mu }_\nu {x^\nu }$, the transformation (\ref{SLor}) can be represented as
\begin{equation}
\label{SLor1}
\begin{gathered}
\psi '(x) = \exp \left( {\frac{i}{2}{\omega _{\mu \nu }}{J^{\mu \nu }}} \right)\psi (x),\\
J^{\mu \nu } = i({x^\mu }{\partial ^\nu } - {x^\nu }{\partial ^\mu }) + \frac{1}{2}{\Sigma ^{\mu \nu }},
\end{gathered}
\end{equation}
where $J^{\mu \nu }$ are complete generators
of the Lorentz group, including generators
$\Sigma^{\mu\nu}$ in the spinor representation.

 Using the complex conjugation (\ref{cdir2}),
  one can construct for the bispinor (\ref{dirspin})
  another conjugated bispinor
\begin{equation}
\label{chc0}
  \psi ^c =\begin{pmatrix}  0&\varepsilon  \\
  {\tilde \varepsilon }&0
\end{pmatrix}\psi ^* = \begin{pmatrix}
  \varepsilon_{ab} \, {\bar \eta }^b \\[0.1cm]
  \varepsilon^{\dot{a}\dot{c}} \, {\bar \xi }_{\dot c}
  \end{pmatrix} =
  \begin{pmatrix}
  {\bar \eta }_a \\[0.1cm]  {\bar \xi }^{\dot a}
\end{pmatrix},
\end{equation}
which is transformed according to the same law (\ref{SLor})
as the initial bispinor $\psi$ given in (\ref{dirspin}).
Here we introduce matrices
\begin{equation}
\label{epsilon}
\varepsilon =||\varepsilon_{ab}|| \, , \;\;\;\;
\tilde{\varepsilon} = ||\varepsilon^{\dot{a}\dot{c}}|| \, , \;\;\;\;
  \varepsilon  =  - \tilde \varepsilon  = -  i{\sigma _2}
  = \begin{pmatrix}  0&- 1 \\   { 1}&0 \end{pmatrix} \, ,
\end{equation}
that lower ${{\bar \eta }_a} = {\varepsilon _{ab}}{{\bar \eta }^b}$
and raise ${{\bar \xi }^{\dot a}} =
{ \varepsilon ^{\dot a\dot b}}{{\bar \xi }_{\dot b}}$
indices of the components of Weyl spinors.
The bispinor $\psi ^c$, defined in (\ref{chc0}),
 is called the {\it charge conjugated to the bispinor}
  $\psi$ (\ref{dirspin}).
  The definition of charge conjugation
  (\ref{chc0}) is written in matrix form as follows:
\begin{equation}
\label{chc}
\psi ^c = C{\bar \psi }^T = C\, (\gamma ^{0})^T \, \psi ^* \, ,
\;\;\;\;  C = \begin{pmatrix}
  \varepsilon &0 \\
  0&{\tilde \varepsilon }
\end{pmatrix} = i{\gamma ^2}{\gamma ^0} \, ,
\end{equation}
and charge conjugation is the involutive operation $(\psi^c)^c=\psi$. The charge conjugation matrix $C$ satisfies the relations
\begin{equation}
\label{ccc}
(\gamma^\mu )^{T} =  - {C^{ - 1}}{\gamma ^\mu } \, C \, ,\quad
{C^T} =  - C \, ,\quad {C^ + } = {C^{ - 1}}\, . \quad C^{- 1} = - C \, .
\end{equation}
The last two relations in (\ref{ccc}) are valid only in specific representations of the Dirac matrices, for example, in the Weyl representation used.
From (\ref{chc}) and (\ref{ccc}) we find the
Dirac conjugated bispinor to $\psi ^c$
\begin{equation}
\label{bpc}
 \bar \psi ^c \equiv \overline{\psi ^c} = \psi^{T}C = - \psi^{T}C^{-1}.
 \end{equation}

 Note that the first relation in (\ref{ccc}) is actually
 the definition of the matrix $C$, since it precisely guarantees that if the bispinor $\psi$ describes a particle with charge $e$, then the bispinor $\psi^c $ describes a particle with charge $- e$,
 i.e. an antiparticle (see, for example, \cite{lan}).
From this relation it also follows that
\begin{equation}
\label{g5}
(\gamma ^{5})^{T} =  {C^{ - 1}}{\gamma ^5 }C.
\end{equation}
Taking into account (\ref{g5}) and
$(\gamma^\mu)^+ = \gamma^0 \, \gamma^{\mu} \, \gamma^0$,
we obtain
$$
(\gamma ^{5})^{*} = - C^{- 1}\,\gamma^0 \, {\gamma ^5 }
 \,\gamma^0 \, C \; ,
 $$
 whence for the left-handed and right-handed
 components of the bispinor $\psi^c $ we have:
\begin{equation}
\label{LcRc}
\begin{gathered}
 \psi_{ L}^{c} := \left( {\psi_{ L} }
 \right)^{ c}  = \frac{1}{2}\left(
{1
 + \gamma ^5 } \right)\psi ^{c}  = \left( {\psi ^{c} } \right)_{R} ,\\
\psi_{R}^{c}  := \left( {\psi _{R} } \right)^{c}  = \frac{1}{2}\left(
{1 - \gamma ^5 }
\right)\psi ^{c}  = \left( {\psi ^{c} } \right)_{L} .
\end{gathered}
\end{equation}

Since the bispinors $\psi$ and $\psi^c$,
as it follows from the comparison of (\ref{dirspin})
and (\ref{chc0}), are equally transformed with respect to
$\rm SL(2,\mathbb{C} )$, they can be equated:
\begin{equation}
\label{maj}
\psi = \psi^c.
\end{equation}
A bispinor $\psi$ equal to its charge-conjugated
 bispinor $\psi^c$ is called a {\it Majorana bispinor}.
 Note that the definition (\ref{maj}) can be generalized
 by including an arbitrary phase factor \cite{moh3,ir2}:
 $\psi^c = e^{i\theta}\psi$, which is sometimes convenient
 (see, for example, (\ref{ev})), but one can always choose
 $\theta = 0$ by accordingly redefining the fermion field $\psi$.

 The condition (\ref{maj}) is not invariant under
$\rm U(1)$-transformations $\psi \to e^{i \alpha} \psi$,
and this means that Majorana fermions are truly neutral
particles that are identical to their antiparticles,
i.e. cannot have conserved additive quantum numbers
associated with $\rm U(1)$-symmetries: electric
charge and any fermion numbers (lepton, baryon, etc.).

The condition (\ref{maj}) is equivalent (see (\ref{dirspin}) and (\ref{chc0})) to the equality of the Weyl spinors that make up the Dirac bispinor: $\xi_a = {\bar \eta}_a$, and
the Majorana bispinor is represented in the form
\begin{equation}
\label{maj1}
{\psi _M} = \begin{pmatrix}
{{\xi _a}}\\
{{{\bar \xi }^{\dot a}}}
\end{pmatrix} \equiv \begin{pmatrix}
{{{\bar \eta }_a}}\\
{{\eta ^{\dot a}}}
\end{pmatrix},\quad {\bar \xi ^{\dot a}} = {\varepsilon ^{\dot a\dot b}}{\left( {{\xi _b}} \right)^*},\quad {\bar \eta _a} = {\varepsilon _{ab}}{\left( {{\eta ^{\dot b}}} \right)^*},
\end{equation}
i.e., it is determined by only one left-handed $\xi_a$,
or right-handed $\eta^{\dot a}$, Weyl spinor.

From (\ref{chc}), (\ref{bpc}), (\ref{maj}) and (\ref{ccc}) we
deduce
\begin{equation}
\label{maj2}
{\bar \psi _M} =  - \psi _M^T{C^{ - 1}} =  \psi _M^T{C} = \left( {{\xi ^a},{{\bar \xi }_{\dot a}}} \right),
\end{equation}
which allows us to write the Majorana mass term
in the Lagrangian as
\begin{equation}
\label{majm}
- {\cal L}_M = \frac{m}{2}{\bar \psi _M}{\psi _M} = \frac{m}{2}\psi _M^{T}C{\psi _M} = \frac{m}{2}\left( {{\xi ^a}{\xi _a} + {{\bar \xi }_{\dot a}}{{\bar \xi }^{\dot a}}} \right) = \frac{m}{2}\left( {{\xi ^a}{\varepsilon _{ab}}{\xi ^b} + {{\bar \xi }_{\dot a}}{\varepsilon ^{\dot a\dot b}}{{\bar \xi }_{\dot b}}} \right),
\end{equation}
where the factor $1/2$ is introduced so that the coefficient $2$ does not appear for mass $m$ in the Dirac equation
 obtained by varying in $\xi$ and $\bar\xi$. Using (\ref{maj1}), Eq. (\ref{majm}) can be rewritten in the equivalent form by making the substitutions $\xi \to \bar\eta$ and $\bar\xi \to \eta$.
We emphasize that the convolutions of the quadratic combinations of the components $\xi$ and $\bar\xi$ with antisymmetric $\varepsilon$-symbols in (\ref{majm}) are nonzero due to the anticommutativity of the components of the fermionic fields $\xi$ and $\bar\xi$. This anticommutativity
also ensures that the electromagnetic
current for Majorana fermions is equal to zero:
\[
\begin{gathered}
{{\bar \psi }_M}{\gamma ^\mu }{\psi _M} = \bar \psi _M^c{\gamma ^\mu }\psi _M^c =  - \psi _M^T{C^{ - 1}}{\gamma ^\mu }C\bar \psi _M^T  \\
= {{\bar \psi }_M}C({\gamma ^{\mu}) ^{T}}{C^{ - 1}}{\psi _M} =  - {{\bar \psi }_M}{\gamma ^\mu }{\psi _M} = 0,
\end{gathered}
\]
where relations (\ref{maj}), (\ref{bpc}) and (\ref{ccc})
 are taken into account.

Note that the kinetic term for the Majorana field is written in the form
\begin{equation}
\label{kinM}
\frac{1}{2}{{\bar \psi }_M}{i\gamma ^\mu }{\partial _\mu }{\psi _M} = \frac{i}{2}\left( {{\xi ^a}{{({\sigma ^\mu })}_{a\dot b}}{\partial _\mu }{{\bar \xi }^{\dot b}}
+ {{\bar \xi }_{\dot a}}{{({{\tilde\sigma }^\mu })}^{\dot ab}}{\partial _\mu }{\xi _b}} \right).
\end{equation}
According to (\ref{majm}) and (\ref{kinM}), the
free Majorana field Lagrangian has the form
\begin{equation}
\label{LMf}
\begin{gathered}
{\cal L}_f = \frac{1}{2}{{\bar \psi }_M}\left({i\gamma ^\mu }{\partial_\mu } - m \right){\psi _M} = \\
 = \frac{1}{2}\xi_{_{(L)}}^{c+}
 \left(i\sigma^{\mu}\partial_{\mu}\xi_{_{(L)}}^c
 - m\xi_{_{(L)}}\right ) + \frac{1}{2}\xi_{_{(L)}}^+
 \left({i{\tilde\sigma}^\mu }{\partial_\mu }
 \xi_{_{(L)}} - m\xi_{_{(L)}}^c \right) ,
\end{gathered}
\end{equation}
where the compact matrix notation is used
(see (\ref{gwei}), (\ref{maj1}) and (\ref{maj2})):
\begin{equation}
\label{majL}
{\psi _M} = \begin{pmatrix}
\xi_{_{(L)}} \\
\xi_{_{(L)}}^c
\end{pmatrix} \, , \quad {\bar \psi} _M =
\left( \xi_{_{(L)}}^{c+},\xi_{_{(L)}}^+ \right)\, ,\quad
\xi_{_{(L)}} = (\xi_a) \, , \quad
\xi_{_{(L)}}^c = i\sigma_2 \, \xi_{_{(L)}}^* =
{
(\varepsilon^{\dot{a}\dot{b}}\,
\bar{\xi}_{\dot{b}})}.
\end{equation}

From (\ref{LMf}) the equation of motion in bispinor form
follows
\[
\left(i\gamma ^\mu {\partial_\mu } - m \right)\psi _M  = 0,
\]
as well as the equivalent pair of equations for Weyl spinors
\begin{equation}
\label{eom}
i{\tilde\sigma}^\mu {\partial_\mu }\xi_{_{(L)}} -
m\xi_{_{(L)}}^c  = 0, \quad
i\sigma^{\mu}\partial_{\mu}\xi_{_{(L)}}^c - m\xi_{_{(L)}} = 0 \, ,
\end{equation}
and the second equation is obtained by complex conjugation
from the first one. For a massless fermion with a given
4-momentum $p^{\mu} = (|\bf p|,\bf p)$
 the bispinor $\psi _M(x) \sim \exp (- ip\cdot x)$ and, as follows from (\ref{eom}) in view of (\ref{sigma}),
 the spinors $\xi_{_{(L)}}(x)$ and $\xi_{_{(L)}}^{c}(x)$
 obey independent equations
\[
{\boldsymbol\sigma}  \cdot {\bf n}\,
{\xi}_{_{(L)}} =
- {\xi}_{_{(L)}}   \, ,\quad\quad
 {\boldsymbol\sigma}  \cdot {\bf n}\,
{\xi}_{_{(L)}}^c =
{\xi}_{_{(L)}}^c \, ,\quad\quad
 {\bf n} = {\bf p}/|\bf  p| \, ,
\]
that is, these spinors are indeed left-handed and
right-handed, respectively (see (\ref{LcRc})).

Now we compare the Majorana mass term (\ref{majm})
with the Dirac one:
\begin{equation}
\label{dirm}
- {\cal L}_D = m_D{\bar\psi}\psi = m_D
\left(\eta_{_{(R)}}^{+}\xi_{_{(L)}}
+ \xi_{_{(L)}}^{+}\eta_{_{(R)}}\right) =
m_D\left({\bar\eta}^{a}\xi_a + {\bar\xi}_{\dot a}\eta^{\dot a}\right)\!,
\end{equation}
where we make use of the
notation (see (\ref{dirspin}),  (\ref{cdir}))
\begin{equation}
\label{2wei}
\psi = \begin{pmatrix}\xi_{_{(L)}} \\
\eta_{_{(R)}}\end{pmatrix}, \quad
\xi_{_{(L)}} = (\xi_a), \quad \eta_{_{(R)}} =(\eta^{\dot a}).
\end{equation}
Thus, the Dirac mass term arises only in the presence
of both left-handed  $\xi_{_{(L)}}$ and right-handed
  $\eta_{_{(R)}}$ Weyl spinors, which
are {\it independent} components of the Dirac bispinor
(in contrast to the Majorana bispinor for which
the mass term is determined either by only left-handed  $\xi_{_{(L)}}$,
 or by only right-handed  $\eta_{_{(R)}}= \xi_{_{(L)}}^{c}$, Weyl components).

\label{sec:app}
\section*{Appendix B. Mass term of general form}

\renewcommand{\theequation}{B.\arabic{equation}}
\setcounter{equation}{0}

Now we consider the general mass term including the Majorana
terms of types $L$ and $R$ and the Dirac term:
\begin{equation}
\label{LMD}
\begin{gathered}
- {\cal L}_{MD} = \frac{1}{2}{m_L}\left( {{{\bar \psi }_L}\psi _L^c
+ \bar \psi _L^c{\psi _L}} \right) +
\frac{1}{2}{m_R}\left( {{{\bar \psi }_R}\psi _R^c
+ \bar \psi _R^c{\psi _R}} \right)\\
+ {m_D}\left( {{{\bar \psi }_L}{\psi _R}
+ {{\bar \psi }_R}{\psi _L}}\right)\!.
\end{gathered}
\end{equation}
Let us introduce, following \cite{chli}, Majorana bispinors
\begin{equation}
\label{2maj}
 \begin{gathered}
\lambda = \psi_L + \psi_L^c = \lambda^c, \quad \rho =
\psi_R + \psi_R^c = \rho^c \, , \\
{\lambda = \begin{pmatrix}\xi_{_{(L)}}
\\ \xi_{_{(L)}}^c\end{pmatrix} \, , \quad
\rho = \begin{pmatrix}\eta_{_{(R)}}^c \\ \eta_{_{(R)}}
\end{pmatrix} }
\end{gathered}
\end{equation}
and, taking into account (\ref{LcRc}), we express in their
terms the bispinors included in (\ref{LMD}):
\begin{equation}
\label{psimaj}
\begin{gathered}
\psi_L = \frac{1}{2}
{(1 - \gamma^5)}\lambda, \quad \psi_L^c = \frac{1}{2}
{(1 + \gamma^5)}\lambda,\\
\psi_R = \frac{1}{2}
{(1 + \gamma^5)}\rho, \quad \psi_R^c = \frac{1}{2}{(1 - \gamma^5)}\rho.
\end{gathered}
\end{equation}
Substituting (\ref{psimaj}) into (\ref{LMD}), we obtain
 a more convenient representation of the total mass term:
\begin{equation}
\label{LMD2}
\begin{gathered}
- {\cal L}_{MD} = \frac{1}{2}{m_L}{\bar \lambda}\lambda +  \frac{1}{2}{m_R}{\bar \rho}\rho
+ \frac{1}{2}{m_D}\!\left({\bar \lambda}\rho + {\bar\rho}\lambda\right)  \\
= \frac{1}{2}({\bar \lambda},{\bar \rho})\!\begin{pmatrix}m_L&m_D\\m_D&m_R\end{pmatrix}\!\!\begin{pmatrix}\lambda\\ \rho\end{pmatrix}.
\end{gathered}
\end{equation}
Diagonalizing the symmetric mass matrix in (\ref{LMD2})
by using the unitary matrix $U$ (see equation (\ref{diag})
above in the main text), we deduce
\begin{equation}
\label{md}
U^{T}\begin{pmatrix}m_L&m_D\\ m_D&m_R\end{pmatrix}U = \begin{pmatrix}m_1&0 \\ 0&m_2\end{pmatrix},
\end{equation}
 where
\begin{equation}
\label{am}
\begin{gathered}
U = \begin{pmatrix} -i \cos\theta&\sin\theta\\i\sin\theta&\cos\theta\end{pmatrix},\quad
 \tan(2\theta) =\frac{2m_D}{m_R - m_L},\\
m_{1,2} = \frac{1}{2}\left[\sqrt{(m_R - m_L)^2
+ 4m_D^2} \mp (m_R + m_L)\right].
\end{gathered}
\end{equation}
As a result, (\ref{LMD2}) takes the form of a mass
term for two Majorana fermions
\begin{equation}
\label{2majd}
\begin{gathered}
- {\cal L}_{MD} = \frac{1}{2}{m_1}{\bar \chi_1}\chi_1 +  \frac{1}{2}{m_2}{\bar \chi_2}\chi_2,\\
\begin{pmatrix}\chi_1\\ \chi_2 \end{pmatrix} = U^{+}\begin{pmatrix}\lambda\\ \rho\end{pmatrix},\\
\chi_1 =i\lambda\cos\theta -i\rho \sin\theta, \quad \chi_2 =\lambda\sin\theta + \rho\cos\theta \, ,
\end{gathered}
\end{equation}
where the Majorana bispinors $\lambda$ and $\rho$
were defined in (\ref{2maj}).
Thus, the general mass term (\ref{LMD}) for the
 bispinor (\ref{2wei}) composed of two {\it independent} Weyl spinors
 is in fact a mass term for two Majorana fermions with different masses.

Let there be the following hierarchy of mass parameters
in (\ref{LMD}):
\begin{equation}
\label{mlh}
m_L \ll m_D \ll m_R.
\end{equation}
Then from (\ref{am}) and (\ref{2majd}) we find
\begin{equation}
\label{mlh1}
\begin{gathered}
\theta \simeq \frac{m_D}{m_R} \ll 1,\\
\chi_1 \simeq i\lambda - i\theta \rho, \quad
\chi_2 \simeq \theta\lambda + \rho,\\
m_1 \simeq \frac{m_D^2}{m_R} - m_L, \quad
m_2 \simeq \frac{m_D^2}{m_R} + m_R.
\end{gathered}
\end{equation}
It follows that under the conditions (\ref{mlh}), the fermion $\chi_2$ turns out to be heavy ($ m_2 \gg m_1$), while the
fermion $\chi_1$ is light. In the case of $m_L = 0$, we arrive at the
{\it seesaw mechanism} (SSM) of the generation
of a small mass due to a large mass,
\[
\begin{gathered}
m_1 \simeq  \frac{m_D^2}{m_R} \ll m_R, \quad m_2 \simeq m_R; \\
\chi_1 \simeq i\lambda = i(\psi_L + \psi_L^c), \quad \chi_2 \simeq \rho = \psi_R + \psi_R^c \; .
\end{gathered}
\]
This mechanism was discussed in the main text (see section
{\bf \ref{sec:seesaw}}) in relation with neutrino physics.

Note also that in the standard SSM the
choice of $m_L = 0$ is made due to
the fact that for the generation of
$m_L \neq 0$  it is required to introduce the
Higgs triplet into the theory, since
 the $L$-type Majorana mass term in (\ref{LMD}) for
 neutrinos carries weak isospin 1. The corresponding Lagrangian of the Yukawa interaction of the left-handed lepton doublets with the Higgs triplet $\Delta$ in the $2\times 2$ matrix representation has the form (see, e.g., \cite{gps})
\begin{equation}
\label{tri}
{\cal L}_{\Delta} = -\frac{1}{2}f_{\alpha\beta}{\bar L}_{\alpha L}^{c}i\tau_2 \Delta L_{\beta L} + \text{H.c.},
\end{equation}
where $f_{\alpha\beta} =f_{\beta\alpha}$ is the matrix of Yukawa couplings and the triplet
\begin{equation}
\label{D}
\Delta = \begin{pmatrix}\Delta^{+}&\sqrt{2}\Delta^{++}\\\sqrt{2}\Delta^{0}&- \Delta^{+}\end{pmatrix}.
\end{equation}
The triplet vacuum expectation value consistent with conservation of electric charge is given by
\begin{equation}
\label{Dv}
\langle{0}|\Delta|{0}\rangle = \begin{pmatrix}0&0\\v_T&0\end{pmatrix}.
\end{equation}
Equations (\ref{tri}) and (\ref{Dv}) lead to
 the Majorana mass term of $L$-type
\[
{\cal L}_{ML} = -  \frac{1}{2}v_{T}f_{\alpha\beta}{\bar \nu}_{\alpha L}^{c}\nu_{\beta L} + \text{H.c.}
\]
In the case of one lepton generation, we obtain
\[
m_L = fv_{T}.
\]
When fixing conditions
\begin{equation}
\label{LR0D}
m_L = m_R = 0,
\end{equation}
we obtain the usual Dirac fermion, which corresponds, as follows from (\ref{am}) and (\ref{2majd}), to the degenerate case of
two Majorana fermions \cite{chli}:
\begin{equation}
\label{D2M}
\begin{gathered}
m_1 = m_2 = m_D,\\
\chi_1 = \frac{i}{\sqrt{2}}(\lambda - \rho), \quad \chi_2 = \frac{1}{\sqrt{2}}(\lambda + \rho).
\end{gathered}
\end{equation}
Taking into account (\ref{2maj}), from (\ref{psimaj}), we find a representation of the Dirac field in the form of a superposition of two Majorana fields with the same mass (see also \cite{chli,bil})
\[
\psi = \frac{1}{{\sqrt2}}(\chi_2 +i\gamma^5 \chi_1).
\]

A small deviation from the case of Dirac fermions leads to quasi-Dirac fermions ({\it quasi-Dirac, or pseudo-Dirac, neutrinos}
in neutrino physics, see \cite{anam}).
Having determined two small parameters
\[
\epsilon = \frac{m_R + m_L}{2m_D} \ll 1, \quad \delta = \frac{m_R - m_L}{4m_D} \ll 1,
\]
and taking into account (\ref{am}) and (\ref{2majd}),
we obtain (cf. (\ref{D2M}))
\[
\begin{gathered}
\chi_1 =\frac{i}{\sqrt{2}}\left[(1 + \delta)\lambda - (1 - \delta)\rho)\right], \quad m_1 = m_D(1 - \epsilon),\\
\quad \chi_2 = \frac{1}{\sqrt{2}}\left[(1-\delta)\lambda + (1 + \delta)\rho)\right], \quad m_2 = m_D(1 + \epsilon).
\end{gathered}
\]

\section*{Appendix C. Diagonalization of the mass matrix
for Majorana fermions}

\renewcommand{\theequation}{C.\arabic{equation}}
\setcounter{equation}{0}

In the Lagrangians describing the seesaw mechanism
(for the case of several generations of neutrinos), the
Majorana mass terms  are written (after sponta\-neous symmetry breaking)
with the help of the complex symmetric mass matrix $M_R$ (see (\ref{SMNR})).
Let the number of the generations of neutrinos be equal to $n$.
To pass to the neutrino states with definite masses,
the symmetric complex $n \times n$ matrix $M_R$
has to be diagonalized
by using the unitary transformation of right-handed
Majorana neutrinos $N_R \to U \, N_R$, where $U \in U(n) $.
This leads to the transformation of the mass matrix
$M_R \to U^T \, M_R \, U$.
The following statement known in the literature
as {\it Takagi's diagonalization theorem} holds
(see Appendix D in \cite{drei} as well as the references therein). \\
 {\bf Theorem.} {\it For any complex symmetric $n\times n$ matrix $M_{\mathbb{C}}$ there exists a unitary matrix $U$ such that
\begin{equation}
\label{matd00}
U^{T}M_{\mathbb{C}}U =  {\rm diag}(m_1,m_2, ... ,m_n)
\equiv M_D \, ,
 \end{equation}
where all parameters $m_j$ are real and non-negative. } \\
{\bf Proof.} The proof is based on the explicit construction
 of the unitary matrix $U$, which diagonalizes
  $M_{\mathbb{C}}$ according to (\ref{matd00}).

The complex symmetric $n\times n$ matrix
$M_{\mathbb{C}}$ is representable
in the form $M_{\mathbb{C}} = X + i Y$, where $X,Y$ are real
symmetric $n\times n$ matrices. Introduce a symmetric real
$2n\times 2n$ matrix that is the $2 \times 2$ block matrix with
  $n\times n$ blocks $X,Y$:
\begin{equation}
\label{mat2n}
M = \left(\begin{array}{cc}
X & - Y \\ [0.1cm]
-Y & -X
\end{array}
\right) = \sigma_3 \otimes X - \sigma_1 \otimes Y
\; , \;\;\;\;\; M^T = M \; ,
\end{equation}
where $\sigma_1,\sigma_3$ are the Pauli matrices
 (\ref{sigma1}). It is known that any symmetric
  real matrix is diagonalized by using
 a real orthogonal matrix $O$:
 \begin{equation}
\label{matd01}
O^T \, M \, O =
{\rm diag}(\lambda_1,\lambda_2,...,\lambda_{2n})\, ,
\;\;\;\;\;\; O^T \, O = I_{2n} \; ,
\end{equation}
where $I_{2n}$ denotes the $2n$-dimensional unit
 matrix.
Obviously, all diagonal elements
$\lambda_k$ are real numbers. Taking into account $(O^T)^{-1} = O$,
relations (\ref{matd01}) are written in the components as
\begin{equation}
\label{matd02}
M_{ik} O_{k \ell} =  O_{i \ell}\, \lambda_\ell \, ,
\;\;\;\;\;\;\;
O_{k i} O_{k \ell} = \delta_{i\ell} \; .
\end{equation}
Now we introduce the set of real $2n$-dimensional
  vectors $v^{(\ell)}$ with coordinates
  $v^{(\ell)}_i = O_{i \ell}$ (i.e., the vectors $v^{(\ell)}$ are
 the columns of the matrix $O$).
From the relations (\ref{matd02}) it becomes clear that
  $v^{(\ell)}$ are eigenvectors of the matrix $M$ with eigenvalues $\lambda_\ell$:
 \begin{equation}
\label{matd03}
M \, v^{(\ell)} =
(\sigma_3 \otimes X - \sigma_1 \otimes Y)\, v^{(\ell)}
= \lambda_\ell \, v^{(\ell)} \; ,
\end{equation}
  where the set of $2n$ vectors $v^{(\ell)}$ forms an orthonormal system:
  $(v^{(\ell)} , v^{(i)}) = \delta_{\ell i}$,
  and therefore defines a basis in $\mathbb{R}^{2n}$.

Now we note that if $\lambda_\ell$ is an eigenvalue of $M$, then $-\lambda_\ell$ is also an eigenvalue of $M$. Indeed, let us multiply both sides of relation (\ref{matd03}) from the left by the non-singular matrix
  $$
  B := i (\sigma_2 \otimes I_n)=
  \begin{pmatrix}0 & I_n\\-I_n & 0 \end{pmatrix}
  $$
 and use the obvious relation
  $B \, M  = - M \, B$. As a result, we obtain
  $M \, (B \, v^{(\ell)}) = -
  \lambda_\ell \, (B \, v^{(\ell)})$, i.e.
  $B \, v^{(\ell)}$ is an eigenvector of the matrix
  $M$ with an eigenvalue $-\lambda_\ell$. Thus, all $2n$ eigenvalues of the matrix $M$ are divided into pairs $(\lambda_\ell,-\lambda_\ell)$, and each pair has one obviously non-negative eigenvalue. We choose all such non-negative eigenvalues\footnote{If both eigenvalues
  in a pair are equal to zero, then either of them is chosen.}, of which there are $n$, and denote these eigenvalues as $m_1,m_2,...m_n$, while
  the corresponding $2n$-dimensional real eigenvectors are
   denoted as $V^{(1)},...,V^{(n)}$:
  \begin{equation}
\label{matd04}
  M \, V^{(k)} = m_k \, V^{(k)} \;\;\;\; \Rightarrow
  \;\;\;\;
  \left(\begin{array}{cc}
X & - Y \\ [0.1cm]
-Y & -X
\end{array}
\right)
\begin{pmatrix}u^{(k)} \\w^{(k)} \end{pmatrix}
= m_k \;
\begin{pmatrix}u^{(k)} \\w^{(k)} \end{pmatrix}
\, ,
  \end{equation}
where $k=1,...,n$,
and we composed $2n$-dimensional vectors $V^{(k)}$
of two $n$-dimensional vectors $u^{(k)}$ and $w^{(k)}$.
Note that the orthonormality property
for any selection of vectors $V^{(k)}$ is preserved, and we have
 \begin{equation}
\label{matd05}
 \delta_{k\ell} =(V^{(k)},V^{(\ell)})
= (u^{(k)},u^{(\ell)}) +
(w^{(k)},w^{(\ell)}) \; .
 \end{equation}
 Let us introduce $n$-dimensional complex vectors
 $z^{(k)} = u^{(k)} + i \, w^{(k)},~
 k=1,...,n$.
 Then the second equality in (\ref{matd04})
 and the orthonormality condition (\ref{matd05}) can be written as
  \begin{equation}
\label{matd06}
M_{\mathbb C} \; z^{(k)} =
m_k \; z^{(k)*} \; , \;\;\;\;\;\;
 (z^{(k)*}, z^{(\ell)}) = \delta_{k\ell} \; ,
\end{equation}
where $z^{(k)*} = u^{(k)} - i \, w^{(k)}$
 and $m_k  \in \mathbb{R}_{\geq 0}$. Now we define the complex
 $n \times n$-matrix $U$, the columns
 of which are the vectors $z^{(k)}$,
 i.e. $U_{ik} = z^{(k)}_i$ $(i ,k=1,...,n)$.
Then the relations (\ref{matd06})
  are presented in the form
 $$
 M_{\mathbb C}  \, U = U^* \, {\rm diag}(m_1,...,m_n)
 \; , \;\;\;\;\;\; U^\dagger \, U = I_n \; .
 $$
Thus, we have constructed a unitary matrix $U \in U(n)$,
which diagonizes, according to (\ref{matd00}), the complex symmetric
$n\times n$ matrix $M_{\mathbb C}$, and,
in addition, the diagonal elements $m_k$
are non-negative real numbers, as required.

\vspace{0.3cm}

{\bf Acknowledgments.} The authors are grateful to A.B.~Arbuzov,
E.E.~Boos and R.~Shrock for useful discussions and comments.

\nocite{*}
\bibliographystyle{pepan}

\end{document}